\documentclass[twocolumn,
aps,
prx,
showpacs,
amsmath,
amssymb,
floatfix,
longbibliography,
superscriptaddress,
eprint]
{revtex4-1}

\usepackage{subfigure}
\usepackage{amsfonts} 

\usepackage{xcolor}
\definecolor{darkblue}{RGB}{0,0,150}
\definecolor{nightblue}{RGB}{0,0,100}

\usepackage{graphicx,mathtools,bm,bbm}
\usepackage{booktabs}
\usepackage{MnSymbol}
\usepackage[
colorlinks,
citecolor=darkblue,
linkcolor=darkblue,
urlcolor=nightblue]{hyperref}
\usepackage{xurl}

\usepackage[english]{babel}
\usepackage[babel,kerning=true,spacing=true]{microtype}
\usepackage[utf8]{inputenc}

\usepackage{soul}

\begin{document}

\title{Surface code off-the-hook: diagonal syndrome-extraction scheduling}
    
\author{Gilad Kishony}
\affiliation{Classiq Technologies. 3 Daniel Frisch Street, Tel Aviv-Yafo, 6473104, Israel.}
\author{Austin Fowler}

\begin{abstract}
In the rotated surface code, hook errors (errors on auxiliary qubits midway through syndrome extraction that propagate to correlated two-qubit data errors) can reduce the circuit-level code distance by a factor of two if the extraction schedule is poorly chosen.
The traditional approach uses N-shaped and Z-shaped schedules, selecting the orientation in each plaquette to avoid hook errors aligned with logical operators. However, this becomes increasingly complex within lattice surgery primitives with varied boundary geometries, and requires a 7-step schedule to avoid gate collisions.
We propose the diagonal schedule, which orients hook errors along the diagonal of each plaquette.
These diagonal errors crucially never align with logical operators regardless of boundary orientation, achieving full code distance.
The diagonal schedule is globally uniform: all X-type plaquettes use one schedule and all Z-type plaquettes use another, eliminating geometry-dependent planning.
On hardware supporting parallel measurement, reset, and gate operations, the schedule achieves a minimal period of 6 time steps, compared to 7 for the traditional approach.
We demonstrate effectiveness for memory experiments, spatial junctions, spatial Hadamard gates, and patch rotation, showing equivalent or improved logical error rates while simplifying circuit construction.
\end{abstract}

\maketitle

\section{Introduction}
\label{sec:intro}

The surface code~\cite{kitaev2003fault, dennis2002topological, fowler2012surface} is a leading candidate for fault-tolerant quantum computation due to its high threshold, local connectivity requirements, and compatibility with planar hardware architectures.
In the rotated surface code~\cite{bombin2007optimal, orourke2025compare}, data qubits lie on the vertices of a square lattice, and quantum information is protected by repeatedly measuring weight-4 stabilizers (X and Z Pauli products on alternating plaquettes).

Each stabilizer measurement couples an auxiliary qubit to the four data qubits of a plaquette via four two-qubit gates and then measures the auxiliary.
The order of these gates (the syndrome-extraction schedule) determines how faults propagate.
A particularly problematic mechanism is a \emph{hook error}~\cite{dennis2002topological, kishony2025increasingdistancetopologicalcodes, liu2026alphasyndrometacklingsyndromemeasurement, viszlai2026prophuntautomatedoptimizationquantum}: an error on the auxiliary qubit midway through the circuit that propagates through the remaining two-qubit gates into a correlated error on two data qubits.
The affected data qubits are those interacting with the auxiliary in the second half of the circuit, and the Pauli type of the resulting two-qubit error matches the Pauli type of the stabilizer being measured.

If the schedule is poorly designed, hook errors can reduce the effective code distance by a factor of two relative to the phenomenological noise case.
This occurs when hook errors are oriented along the direction of logical operators, i.e., paths connecting boundaries of the same Pauli type.

The standard mitigation uses two schedule variants, commonly called N-shaped and Z-shaped schedules based on the ordering pattern around the plaquette~\cite{tomita2014surface17, google2023suppressing, geher2024hadamard}.
The N-shaped schedule produces vertical hook errors, while the Z-shaped schedule produces horizontal hook errors.
By choosing between these schedules per plaquette according to the geometry of nearby boundaries, one can (in most cases) orient hook errors perpendicular to the problematic logical-operator directions.

\begin{figure}
    \centering
    \includegraphics[width=\columnwidth]{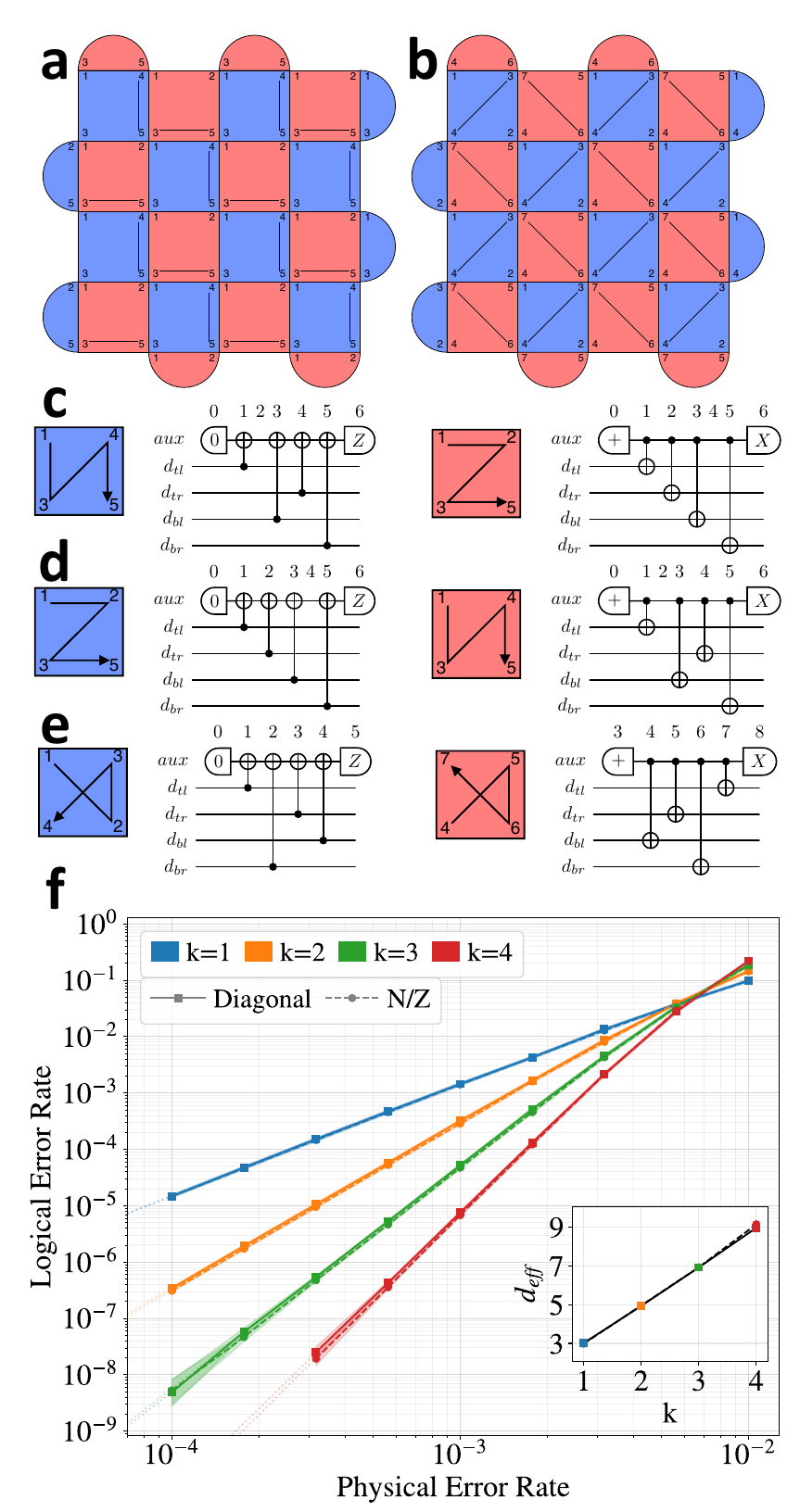}
    \caption{
    \textbf{Memory experiment} comparison.
    \textbf{(a)} Traditional N/Z schedule for a memory experiment, with schedule orientations chosen based on boundary types (hook error orientations shown).
    \textbf{(b)} Diagonal schedule for the same memory experiment, using a globally uniform schedule (hook error orientations shown).
    \textbf{(c)} Z-shaped X-plaquette and N-shaped Z-plaquette circuits with their corresponding schedule diagrams.
    \textbf{(d)} Alternative orientation: N-shaped X-plaquette and Z-shaped Z-plaquette.
    \textbf{(e)} Diagonal X-plaquette and Z-plaquette circuits.
    \textbf{(f)} Logical error rate versus physical error rate for standard (dashed) and diagonal (solid) schedules at various code distances $d = 2k+1$.
    The two approaches achieve nearly identical logical error rates.
    Inset: effective distance $d_{\mathrm{eff}}$ (extracted by fitting the slope of $p_{\mathrm{logical}}$ vs.\ $p$ at low $p$) versus $k$, confirming $d_{\mathrm{eff}} \approx d = 2k+1$ for both schedules.
    }
    \label{fig:memory}
\end{figure}

For a simple memory experiment with a single patch (e.g., X boundaries on the top and bottom and Z boundaries on the sides), the choice is straightforward: use Z-shaped schedules for X stabilizers (horizontal hooks) and N-shaped schedules for Z stabilizers (vertical hooks).
This prevents hook errors from creating short paths between same-type boundaries (Fig.~\ref{fig:memory}).

When performing computation using lattice surgery~\cite{horsman2012lattice,litinski2019game}, the situation becomes significantly more cumbersome.
Lattice surgery operations create intricate three-dimensional spacetime geometries with boundaries of different Pauli types in various directions, making schedule selection per plaquette and per time step difficult and potentially impossible in some cases.
In addition, accommodating different N/Z orientations in adjacent regions requires a 7-step schedule to prevent collisions where neighboring plaquettes would simultaneously apply gates to a shared data qubit.

An alternative is to alternate the gate ordering between consecutive rounds of syndrome extraction~\cite{bluvstein2025architectural, gidney2024alternating}.
This allows using \emph{any} schedule on each plaquette (even a ``bad'' choice producing hook errors aligned with logical operators) by alternating it with its time-reversed version each round.
With this temporal alternation, only every other round has unfavorable hook propagation, and the effective code distance becomes $d-1$ rather than the $\lceil d/2 \rceil$ that would result from using a fixed bad schedule.
However, sacrificing even a single unit of distance is costly, especially at the small distances relevant to near-term experiments.

\section{The Diagonal Schedule}
\label{sec:diagonal}

We propose the diagonal schedule to avoid geometry-dependent N/Z planning while preserving full distance.
In the diagonal schedule, the four two-qubit gates are ordered so that the first two act on one diagonal pair of data qubits of the plaquette and the last two act on the opposite diagonal pair (Fig.~\ref{fig:memory}).

Hook errors in this schedule therefore affect two data qubits along a plaquette diagonal.
Crucially, \emph{diagonal hook errors never align with the direction of any logical operator}, regardless of boundary geometry.
Logical operators run between boundaries of the same type along paths on the square lattice, and a diagonal two-qubit error (neither purely horizontal nor purely vertical) cannot create a shortcut along such a path.
As a result, the diagonal schedule can be used uniformly in space-time without geometry-dependent planning: all X-type plaquettes use one schedule and all Z-type plaquettes use another.

In order for a stabilizer measurement schedule to be valid (for the parity of the intended stabilizer to be measured through the auxiliary measurement), it must satisfy the following constraint~\cite{beverland2021cost}:
whenever two stabilizers of opposite Pauli types share data qubits, an even number of the shared data qubits must be coupled to one auxiliary (via two-qubit gates) before they are coupled to the other auxiliary.
In the surface code, neighboring stabilizers of different types overlap on two data qubits; in these cases, both shared data qubits must be coupled to the auxiliary of one stabilizer before either is coupled to the auxiliary of the other.
All N/Z schedules shown in Fig.~\ref{fig:memory} satisfy this constraint regardless of how they are tiled, and the diagonal schedule in Fig.~\ref{fig:memory}(e) is designed to satisfy it as well.

Fig.~\ref{fig:memory} shows one cycle of syndrome extraction for bulk X and Z plaquettes under each schedule [N/Z orientations in panels (c,d), diagonal in panel (e)].
Each cycle consists of auxiliary reset, four two-qubit gates, and auxiliary measurement.
In the N/Z schedules, the auxiliary is idle for one time step; this idling supports mixing schedule orientations across the lattice while avoiding gate collisions on shared data qubits.
In contrast, the diagonal schedule is globally uniform, so auxiliary qubits are never idle.

The period (time steps per syndrome extraction cycle) depends on both the schedule choice and hardware capabilities.
When measurements and resets must be performed in separate steps from entangling gates, the N/Z schedule has period 7, with reset at time step 0 (mod 7) and measurement at time step 6 (mod 7).
If the N/Z schedule is used uniformly in space, its period can be reduced to 6 by removing the idling step, as done in Fig.~\ref{fig:memory}(f).
The diagonal schedule in Fig.~\ref{fig:memory}(e) has period 9, with reset at time step 0 (mod 9) and measurement at time step 8 (mod 9).
The diagonal X-plaquette schedule can be shifted earlier by one time step without introducing gate collisions; this delay is included only to facilitate the spatial Hadamard construction (Section~\ref{sec:hadamard}), and in time slices without spatial Hadamard operations the period can be reduced to 8 by removing it.

When measurements and resets can execute in parallel with entangling gates, the N/Z schedule remains at period 7 (or 6 if uniform), while the diagonal schedule achieves period 6: plaquettes start a new cycle immediately after completing the previous one with no idling.
The diagonal schedule is particularly advantageous on hardware where measurements and resets can execute in parallel with gates, or where measurement/reset duration dominates such that additional gate layers have minimal impact on cycle time.
In any case, it should be noted that on most hardware, idling noise is insignificant compared with gate infidelity and measurement/reset errors.
Table~\ref{tab:schedule_comparison} summarizes these tradeoffs.

\begin{table}[h]
\centering
\begin{tabular}{c|c|c|c}
 & N/Z & Alternating & Diagonal \\
\hline\hline
Period & 7 & 6 & 6 \\
(parallel M/R) & & & \\
\hline
Period & 7 & 6 & 8--9 \\
(sequential M/R) & & & \\
\hline
Distance & $\leq d$ & $d-1$ & $d$ \\
\hline
Construction & geometry- & any schedule & globally \\
 & dependent & + reversal & uniform \\
 & (complex) & (simple) & (simple)
\end{tabular}
\caption{Comparison of syndrome extraction scheduling approaches.
\textbf{N/Z}: N-shaped or Z-shaped gate orderings, orientation chosen per plaquette based on boundary geometry.
\textbf{Alternating}: any base schedule, alternated with its time-reversal between consecutive rounds.
\textbf{Diagonal}: same diagonal gate ordering for all plaquettes of each type.
``Parallel M/R'': measurements/resets execute in parallel with gates; ``sequential M/R'': they occur in separate time steps.}
\label{tab:schedule_comparison}
\end{table}

\section{Simulations}
\label{sec:simulations}

In the following sections, we benchmark the diagonal schedule in increasingly complex lattice surgery settings.
In all cases, we verify using integer linear programming~\cite{landahl2011faulttolerant} that the \emph{diagonal schedule achieves the expected maximal circuit-level code distance}.

Throughout this work, we use a uniform depolarizing noise model: each gate (including idling) is followed by depolarizing noise at physical error rate $p$, each reset is flipped with probability $p$, and each measurement outcome is flipped with probability $p$.
We simulate hardware where measurements and resets can be performed in parallel with entangling gates, using the shortest possible period in each case.
We parameterize $d = 2k + 1$, and all circuit diagrams drawn use $k = 2$ (i.e., $d = 5$).
Each simulation runs $d$ rounds of syndrome extraction ($d$ rounds per time step for the block rotation circuit in Section~\ref{sec:rotation}), and we report logical error rates per shot.
In each case, we initialize and measure in the basis corresponding to the logical operator with the smallest distance.

To extract the effective distance $d_{\mathrm{eff}}$ from the simulation data, we fit the logical error rate to $p_{\mathrm{logical}} \propto p^{t+1}$ at small $p$, where $t = \lfloor (d-1)/2 \rfloor$ is the number of adversarial errors the code can correct.
To leading order in $p$, the logical error rate is dominated by errors of weight $t+1$.
We define $d_{\mathrm{eff}} = 2t + 1$ (always odd), expecting $d_{\mathrm{eff}} = 2\lfloor (d+1)/2 \rfloor - 1$ for an optimal decoder; observed $d_{\mathrm{eff}}$ may be lower due to decoder limitations.

All circuits were constructed using the TQEC library~\cite{tqec2024} using the uniform bulk convention in which all patches share the same checkerboard pattern of plaquettes in the bulk.
Circuits were simulated using Stim~\cite{gidney2021stim} and decoding was performed using PyMatching~\cite{higgott2023pymatching} or the Tesseract decoder~\cite{tesseract2024}.
A diagonal hook error triggers four neighboring stabilizer measurements rather than two in the case of vertical or horizontal hook errors, making diagonal hook errors non-matchable.
However, a matching decoder can still be applied by decomposing these correlated two-qubit errors into pairs of independent single-qubit errors, analogous to decomposing a Y error into independent X and Z errors.
This decomposition does not reduce the effective distance achieved by the decoder because diagonal hook errors never form shortcuts for logical operators.
The circuits and Python code used to generate the results are available at \url{https://github.com/kishonyWIS/diagonal_syndrome_extraction_surface_code}.

\section{Memory Experiment}
\label{sec:memory}

We first demonstrate the diagonal schedule in the simplest setting: a memory experiment on a single surface code patch.
Fig.~\ref{fig:memory} compares the traditional and diagonal schedules in this geometry. Panels (a) and (b) show the plaquette layouts for standard and diagonal schedules respectively.
Panel (f) presents a comparison of the logical error rates in the two cases.
Both schedules are simulated with the same period-6 timing which is possible in both cases since the schedules are spatially uniform.
The logical error rates are nearly identical across all physical error rates and all code distances tested.
In the next sections we focus on lattice-surgery geometries where N/Z scheduling becomes cumbersome and can incur a longer period.

\section{Spatial Junctions}
\label{sec:junctions}

Lattice surgery operations require joining and splitting surface code patches, creating complex geometries in 2+1 dimensions.
A surface code patch may be joined with any subset of its four spatially neighboring patches in a particular time slice, forming various spatial junctions.
These junctions present a challenge for traditional N/Z schedule because the optimal hook error orientations vary from location to location.

\begin{figure}
\centering
\includegraphics[width=\columnwidth]{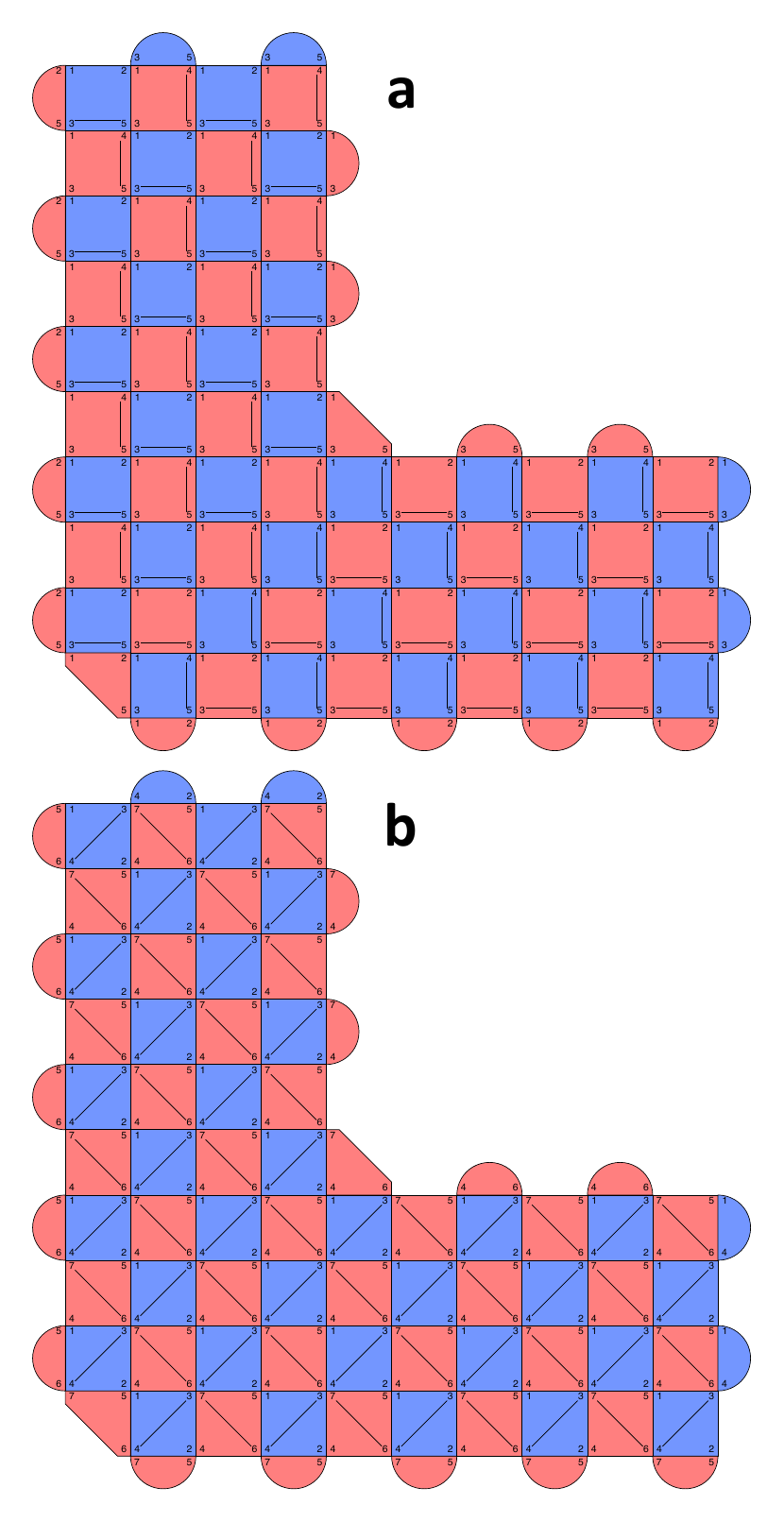}
\caption{
\textbf{L-shaped spatial junction.}
\textbf{(a)} Standard schedule with orientations chosen for the junction geometry.
\textbf{(b)} Diagonal schedule using the globally uniform circuit.
The diagonal schedule simplifies circuit construction while maintaining code distance.
}
\label{fig:l_junction}
\end{figure}

Fig.~\ref{fig:l_junction} shows an L-shaped spatial junction. The L-junction demonstrates how the diagonal schedule eliminates the need to reason about which schedule orientations are appropriate at the junction.
With traditional N/Z scheduling, the plaquettes near the corner require careful analysis to avoid creating hook error paths that could reduce the effective distance.
The diagonal schedule simply uses the same circuit for all plaquettes, automatically avoiding problematic configurations and achieving full code distance.

\begin{figure}
\centering
\includegraphics[width=\columnwidth]{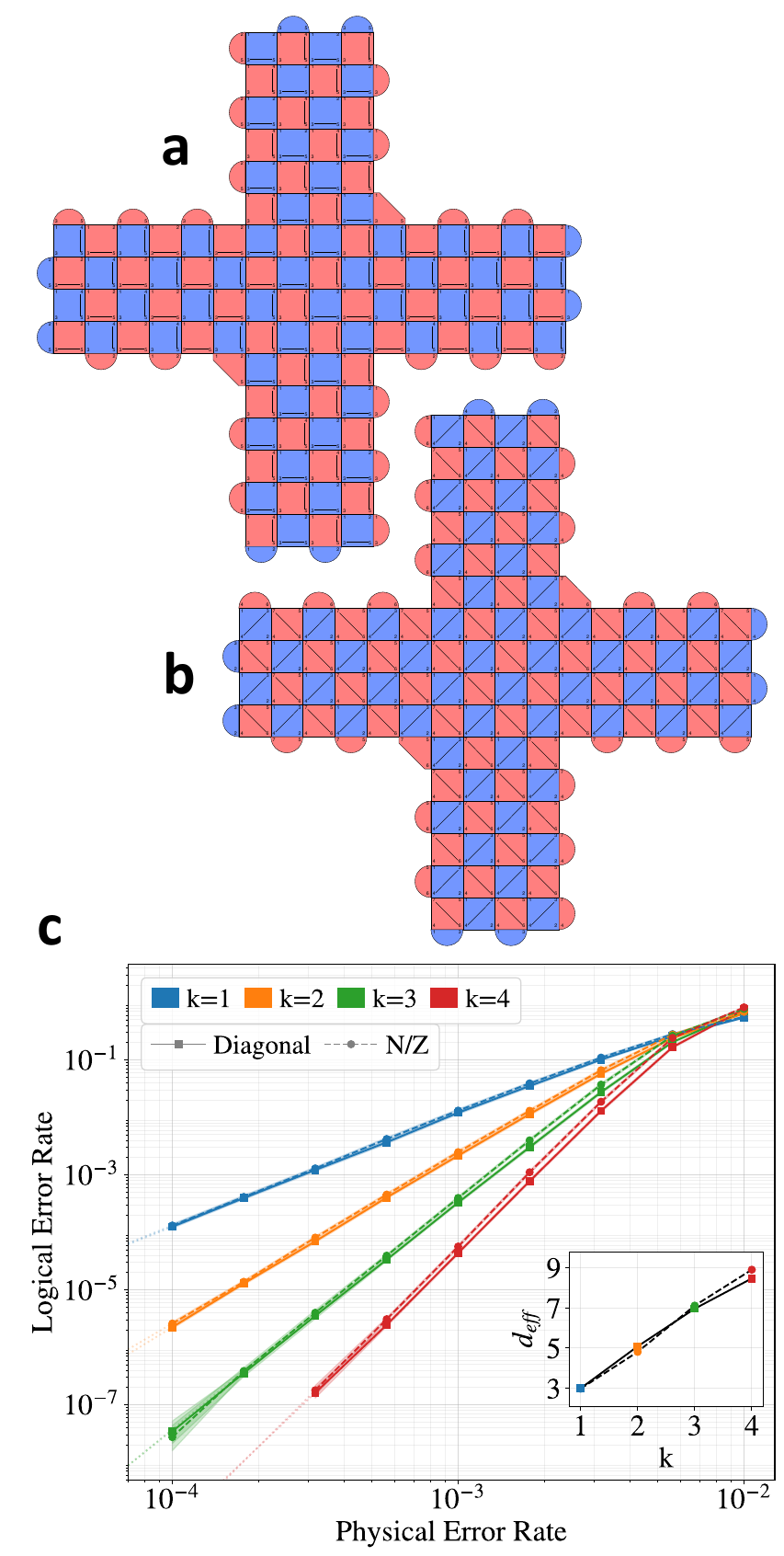}
\caption{
\textbf{X-shaped spatial junction.}
\textbf{(a)} Standard schedule requiring careful orientation planning at the four-way junction.
\textbf{(b)} Diagonal schedule with uniform circuits throughout.
\textbf{(c)} Logical error rate comparison showing the diagonal schedule (solid) slightly outperforming the standard schedule (dashed).
This improvement arises because the diagonal schedule enables a more compact period-6 circuit, while the standard schedule requires period-7 to avoid gate collisions at the junction.
Inset: effective distance $d_{\mathrm{eff}}$ versus $k$, confirming $d_{\mathrm{eff}} \approx d = 2k+1$ for both schedules.
}
\label{fig:x_junction}
\end{figure}

Fig.~\ref{fig:x_junction} shows an X-shaped junction where a central patch is joined to four neighboring patches, representing a more complex lattice surgery configuration.
The X-junction is particularly challenging for traditional scheduling because all of the nearby boundaries are of the same Pauli type, making it difficult to find an orientation that avoids all problematic hook error paths.
Panel (c) shows that the diagonal schedule actually achieves slightly lower logical error rates than the N/Z schedule for this geometry.
This improvement comes from the reduced circuit period: the diagonal schedule's global uniformity allows a period-6 circuit, while the standard schedule requires period-7 to accommodate different orientations in adjacent plaquettes without gate collisions.

\section{Spatial Hadamard}
\label{sec:hadamard}

\begin{figure}
\centering
\includegraphics[width=0.98\columnwidth]{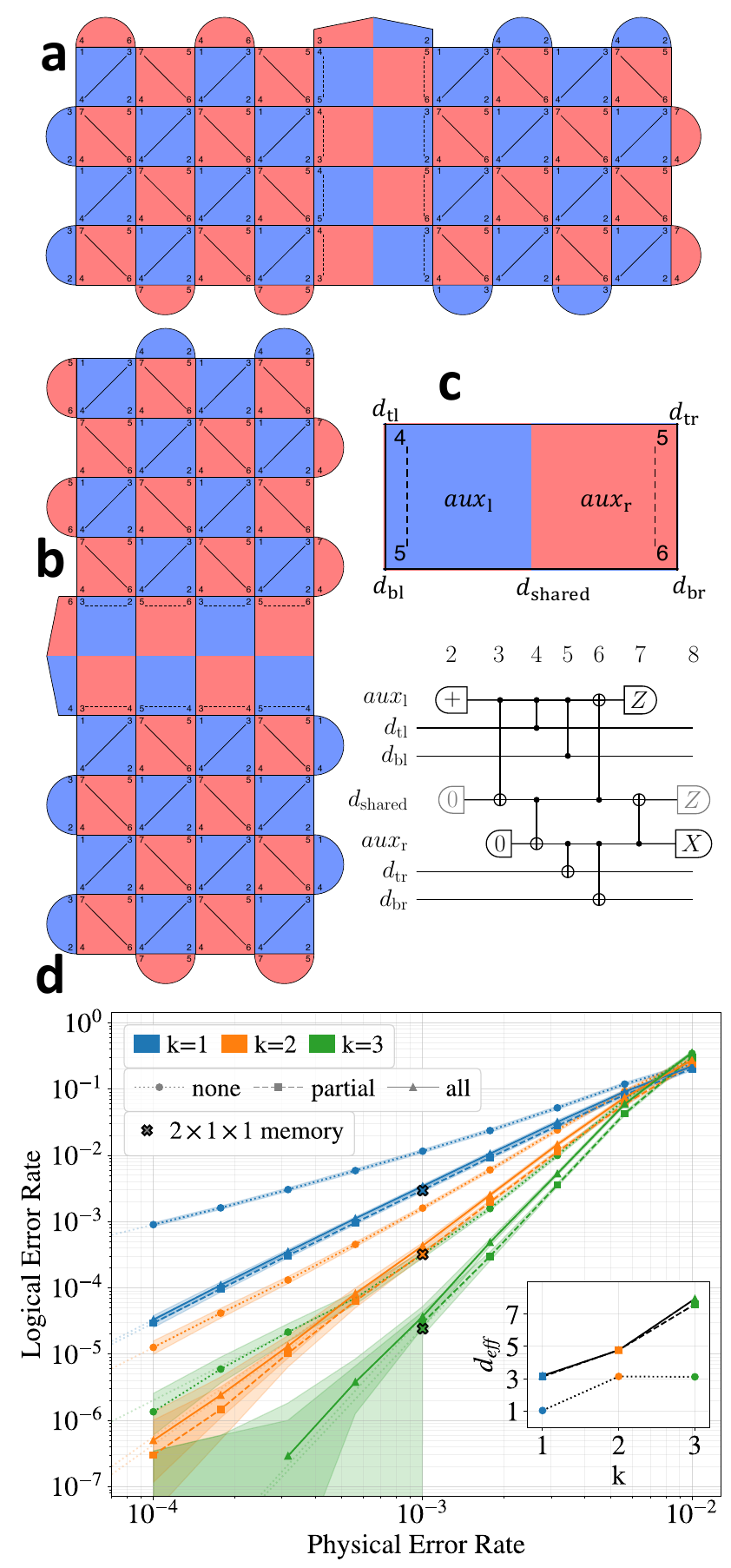}
\caption{
\textbf{Spatial Hadamard} gate with the diagonal schedule.
\textbf{(a)} Vertical orientation.
\textbf{(b)} Horizontal orientation.
\textbf{(c)} Detailed view of one of the stretched stabilizers and the corresponding syndrome extraction circuit.
Circuits for the other stretched stabilizer are similar.
Stretched stabilizers are prone to hook errors along their short direction.
Flag measurements (shown in circuit) detect these problematic hook errors.
\textbf{(d)} Logical error rate versus physical error rate using the Tesseract decoder.
Different flag measurement settings are tested: ``none'' (dotted lines), ``partial'' (dashed lines), and ``all'' (solid lines).
X-shaped markers at physical error rate $10^{-3}$ indicate the logical error rates achieved by the Tesseract decoder for a memory experiment occupying the same volume ($2\times 1\times 1$).
With partial flags, the spatial Hadamard matches these logical error rates.
Inset: effective distance $d_{\mathrm{eff}}$ versus $k$.
The expected effective distances are achieved depending on the flag configuration.
}
\label{fig:hadamard}
\end{figure}

The spatial Hadamard gate~\cite{geher2024hadamard} exchanges the X and Z logical operators of a surface code patch via surgery between a Pauli X boundary of one patch and a Pauli Z boundary of its neighbor [Fig.~\ref{fig:hadamard}(a,b)].
To maintain the uniform-bulk convention, this surgery introduces \emph{stretched stabilizers} along the interface.

The stretched stabilizers span a rectangle composed of two square plaquettes.
Assuming only local connectivity on the square lattice, measuring such a stretched stabilizer requires teleporting partial syndrome information between the two sides of the rectangle through a sequence of entangling gates along a chain of three qubits within the rectangle.
These circuits are therefore particularly susceptible to hook errors along the short axis of the rectangle.
Such hook errors can create logical-error shortcuts if not properly handled.
It should be noted, however, that this distance reduction can possibly be tolerated since it occurs with respect to errors along a particular two-dimensional surface in space-time rather than as a bulk phenomenon.

Nonetheless, we avoid this distance reduction with no extra cost in circuit depth by adding flag measurements~\cite{chao2018flag, gidney2023colorcode} to the stretched stabilizer circuit.
As shown in Fig.~\ref{fig:hadamard}(c), the circuit uses three auxiliary qubits, $aux_{\mathrm{l}}$, $d_{\mathrm{shared}}$, and $aux_{\mathrm{r}}$, which together measure the stabilizer supported on the four corners of the rectangle.
The stabilizer measurement outcome is obtained from the X-basis measurement of $aux_{\mathrm{r}}$.
The other two qubits, $aux_{\mathrm{l}}$ and $d_{\mathrm{shared}}$, can be measured in the Z basis and serve as flag measurements that detect hook errors: specifically, errors on $aux_{\mathrm{l}}$ or $aux_{\mathrm{r}}$ at intermediate times during the circuit.
Flag measurements of $aux_{\mathrm{l}}$ come ``for free'' in circuit depth because they can be performed in parallel with other operations, whereas flag measurements of $d_{\mathrm{shared}}$ require an additional time step.

We test three flag-measurement settings: ``none'' (no flags), ``all'' (measuring both $aux_{\mathrm{l}}$ and $d_{\mathrm{shared}}$), and ``partial'' (measuring $aux_{\mathrm{l}}$ every round and measuring both only after the final round).
Without flag measurements, hook errors on stretched stabilizers reduce the circuit-level code distance by a factor of two (from $2k+1$ to $k+1$).
With either partial or full flag measurements, the circuit achieves full code distance.
A hook error on $aux_{\mathrm{l}}$ is immediately detected by measuring $aux_{\mathrm{l}}$ in the same round, while a hook error on $aux_{\mathrm{r}}$ is detected by measuring $d_{\mathrm{shared}}$ in the same round in the ``all'' setting.
In the ``partial'' setting, a hook error on $aux_{\mathrm{r}}$ propagates to $d_{\mathrm{shared}}$ at the start of the next round and then from there to the next flag measurement of $aux_{\mathrm{l}}$.
The partial setting is therefore an attractive tradeoff: it maintains full code distance while preserving the period-6 cycle of the diagonal schedule.

When flag measurements are performed, these hook-error patterns become non-matchable, and a vanilla matching-based decoder cannot effectively exploit the flag information.
Unlike the bulk diagonal hook errors discussed earlier, which do not reduce the effective distance achieved by the decoder when decomposed into independent errors, the stretched-stabilizer hook errors \emph{do} create shortcuts for logical operators.
For the decoder to restore full distance, it must account for the correlations between the flag measurements and the hook errors they detect; without this correlation information, the decoder effectively operates with respect to a reduced-distance error model (distance $k+1$ rather than $2k+1$), even when flags are measured.
As shown in Fig.~\ref{fig:hadamard}(d), the Tesseract decoder achieves the full effective distance $d_{\mathrm{eff}} = 2k+1$ with ``partial'' and ``all'' flags, proving that the circuit itself achieves full distance.
Furthermore, the logical error rates achieved using ``partial'' flags match those of a memory experiment occupying the same volume ($2\times 1\times 1$).
However, matching-based decoders face limitations: as demonstrated in Appendix~\ref{app:spatial_hadamard_decoder_comparison}, both PyMatching and correlated PyMatching~\cite{fowler2013optimalcomplexitycorrectioncorrelated} achieve only $d_{\mathrm{eff}} = 2\lfloor k/2 \rfloor + 1$ with all flag configurations, despite the circuit achieving full distance.

As shown in Appendix~\ref{app:decoder_runtime}, the Tesseract decoder is $10^3$--$10^4$ times slower than matching-based decoders for these circuits.
This motivates further research into decoders that can achieve the full effective distance for these circuits with a runtime comparable to matching-based decoders.
While the spatial Hadamard serves as a useful lattice-surgery primitive, it can always be replaced with its temporal counterpart, the temporal Hadamard (which does not suffer from the stretched stabilizer complication).
It remains unknown whether this replacement can be achieved within an arbitrary logical computation without increasing its volume.
In any case, the stretched stabilizer complication is a consequence of the uniform-bulk convention rather than the scheduling choice, and its relative impact is expected to diminish with increasing code distance.

\section{Patch Rotation}
\label{sec:rotation}

\begin{figure}
\centering
\includegraphics[width=\columnwidth]{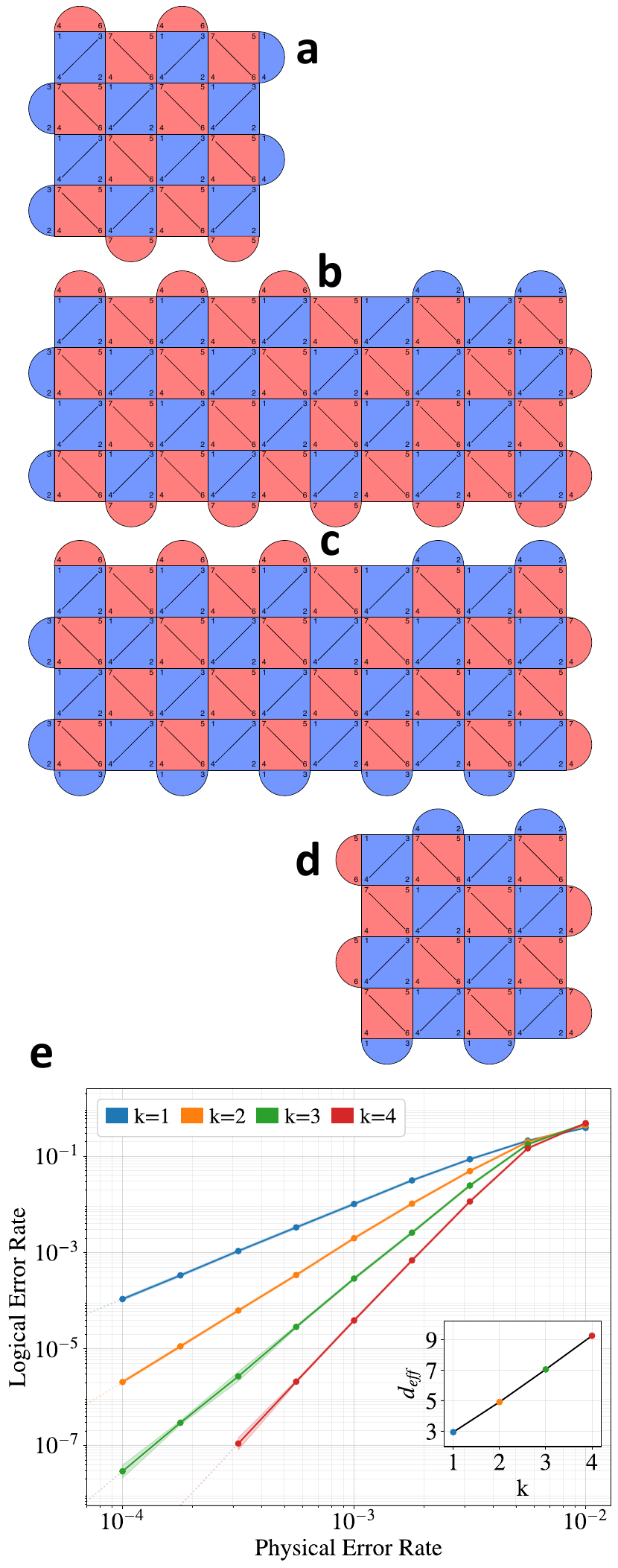}
\caption{
\textbf{Patch rotation} with the diagonal schedule.
\textbf{(a)--(d)} Four time steps during the patch rotation circuit, showing how the patch geometry evolves.
The diagonal schedule maintains its uniform structure throughout the rotation.
\textbf{(e)} Logical error rate versus physical error rate. Different colors indicate different code distances $d = 2k+1$.
Inset: effective distance $d_{\mathrm{eff}}$ versus $k$, confirming $d_{\mathrm{eff}} \approx d = 2k+1$.
}
\label{fig:rotation}
\end{figure}

Patch rotation~\cite{litinski2019game} reorients a surface code patch by $90^\circ$ through a sequence of deformations that temporarily stretch the patch into a two-by-one rectangle and then shift its boundaries before shrinking back to the original size (Fig.~\ref{fig:rotation}(a--d)).
Throughout this process, the diagonal schedule can be applied without modifying its spatially and temporally uniform structure, and we observe no distance reduction despite the evolving boundary geometry.

Figure~\ref{fig:rotation}(e) shows the logical error rate as a function of physical error rate and code distance.
The logical error rate is roughly six times that of the memory experiment, consistent with the fact that the patch-rotation sequence [including the initial and final logical idling steps in panels (a) and (d)] comprises six patches$\times$time steps.

\section{Summary}
\label{sec:summary}

We have presented the diagonal schedule for syndrome extraction in the rotated surface code, which avoids the geometry-dependent orientation planning required by traditional N/Z scheduling while preserving full code distance.
The diagonal schedule offers three practical advantages: \textbf{hook error immunity}, \textbf{simplified construction} (a globally uniform circuit for all plaquettes of a given type), and a \textbf{compact period} especially on hardware supporting parallel measurement/reset with gates.
We demonstrated the diagonal schedule in a memory experiment and across several lattice-surgery primitives (L-junctions, X-junctions, spatial Hadamard gates, and patch rotation), observing equivalent or improved logical error rates while significantly simplifying circuit construction.

For the spatial Hadamard gate, the uniform-bulk convention introduces stretched stabilizers that are susceptible to short-axis hook errors.
We introduced flag measurements that detect these problematic hook errors and maintain full circuit-level distance without increasing circuit depth.
While the Tesseract decoder achieves the full circuit-level distance, matching-based decoders achieve only $d_{\mathrm{eff}} = 2\lfloor k/2 \rfloor + 1$, though this distance reduction is limited to the two-dimensional interface surface.
The correlated matching decoder improves logical error rates relative to vanilla matching, but both face the same asymptotic limitation.
Developing faster decoders that can achieve full effective distance for flagged circuits, or similarly handle heralded erasure errors relevant to certain hardware \cite{jacoby2025magicstateinjectionerasure}, is a promising direction for future work.

Two additional cases expected to benefit from the diagonal schedule are the yoked surface code~\cite{gidney2025yoked} and the crosshairs surface code~\cite{litinski2025crosshairs}.
Both constructions store logical qubits more compactly for a given code distance and involve highly complex boundary patterns, making them natural candidates for a globally uniform schedule with geometry-independent hook-error handling.

More broadly, it would be valuable to investigate syndrome extraction schedules for other quantum error correcting codes, with the goal of avoiding hook-error-induced distance reduction while maintaining low circuit depth.
For example, in the color code~\cite{bombin2006color, beverland2021cost, lee2025colorcode, gidney2023colorcode} it may prove beneficial to consider schedules that are not uniform across all plaquettes.
This raises a fundamental algorithmic question: given an arbitrary (qLDPC~\cite{tillich2014qldpc, panteleev2022qldpc, bravyi2024qldpc}) code, is there an efficient algorithm to choose stabilizer measurement schedules that maximize the circuit-level code distance?


\acknowledgments

We thank Ron Cohen, Shoham Jacoby, Peter-Jan H. S. Derks, Craig Gidney, Oscar Higgott, Adrien Suau, Yaar Vituri, Arabella Schelpe    for useful discussions.

\appendix

\section{Decoder Comparison for Spatial Hadamard}
\label{app:spatial_hadamard_decoder_comparison}

\subsection{Matching Decoders with Full Noise}
\label{app:spatial_hadamard_pure_matching}

\begin{figure}[h]
    \centering
    \includegraphics[width=\columnwidth]{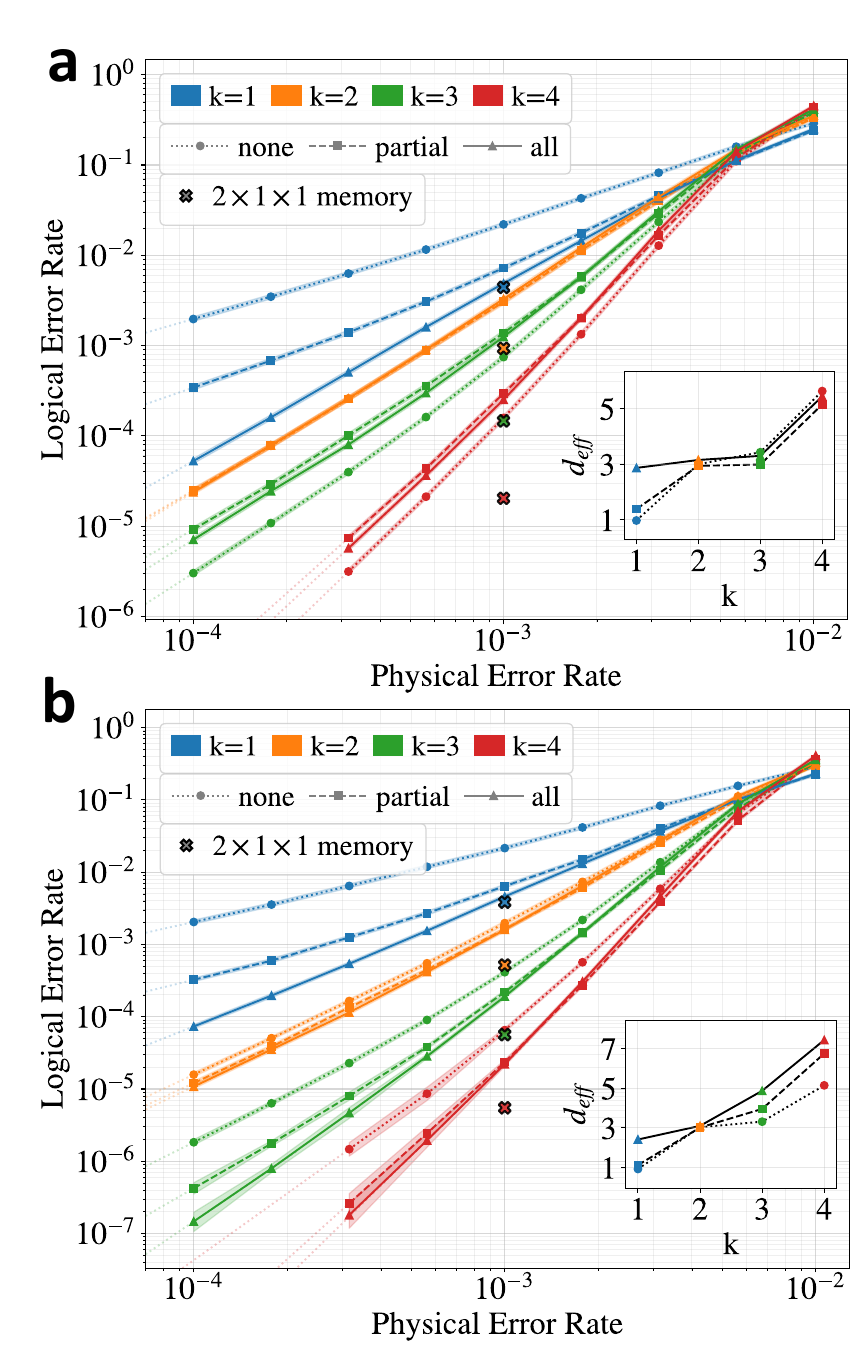}
    \caption{Logical error rates for the \textbf{spatial Hadamard circuit with full noise}, decoded using matching-based decoders.
    \textbf{(a)} PyMatching.
    \textbf{(b)} Correlated PyMatching.
    Different flag measurement settings are tested: ``none'' (dotted lines), ``partial'' (dashed lines), and ``all'' (solid lines).
    X-shaped markers at physical error rate $10^{-3}$ indicate the logical error rates of the $2\times 1\times 1$ memory experiment for reference (using the same decoder as in each panel).
    We are not able to match these memory logical error rates for the Hadamard with these decoders; effective distances are hard to determine from these plots.
    }
    \label{fig:spatial_hadamard_matching_full}
\end{figure}

Fig.~\ref{fig:spatial_hadamard_matching_full} shows the performance of matching-based decoders applied to the spatial Hadamard circuit.
Each panel includes x-shaped markers at physical error rate $10^{-3}$ indicating the logical error rates achieved for a $2\times 1\times 1$ volume memory experiment (occupying the same volume as that simulated for the Hadamard circuit), using the same decoder as used for the spatial Hadamard circuit in that panel.
We are not able to match these memory logical error rates for the Hadamard with these decoders: for PyMatching, the logical error rate achieved for the Hadamard at $k=4$ is roughly 7 times higher than that of the memory experiment; for correlated PyMatching, this ratio is reduced to roughly 4 but remains substantial.
From these plots it is hard to determine the effective distances since the logical error rates are impacted by both bulk errors and hook errors.

\subsection{Interface-Only Noise Model}
\label{app:spatial_hadamard_interface_only}

To clearly reveal the effective distance achieved by different decoders, we use a reduced noise model: errors are applied only to qubits at the stretched stabilizer interface, rather than to all qubits in the circuit.
This isolates the interface behavior and allows measurement of effective distance without requiring very low physical error rates that would necessitate the simulation of prohibitively many shots.
With the full noise model, bulk errors dominate at higher error rates, making it difficult to extract the asymptotic effective distance from the stretched-stabilizer contribution alone.

\begin{figure}[h]
    \centering
    \includegraphics[width=\columnwidth]{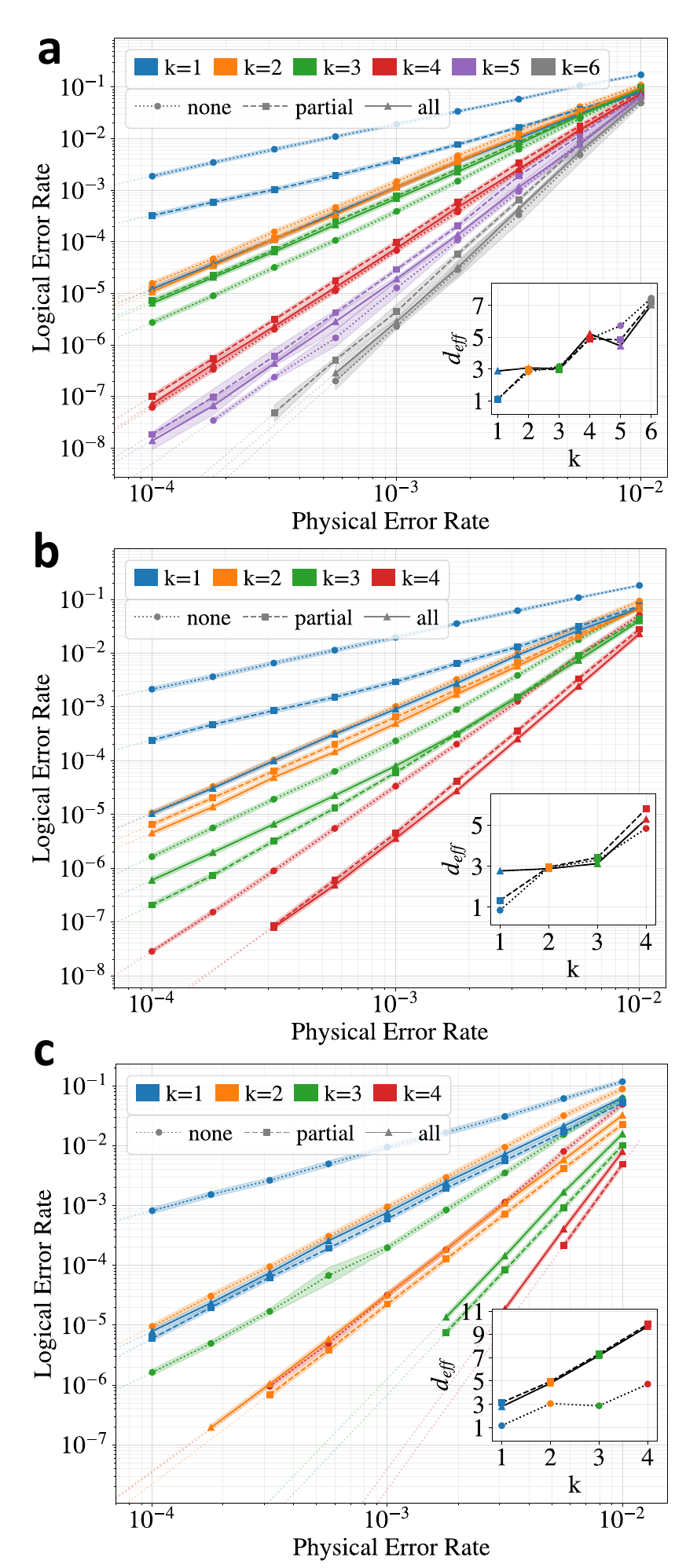}
    \caption{Logical error rates for the \textbf{spatial Hadamard circuit with noise only at the interface}.
    \textbf{(a)} PyMatching.
    \textbf{(b)} Correlated PyMatching.
    \textbf{(c)} Tesseract.
    Different flag measurement settings are tested: ``none'' (dotted lines), ``partial'' (dashed lines), and ``all'' (solid lines).
    Insets: effective distance $d_{\mathrm{eff}}$ versus $k$ for each decoder.
    }
    \label{fig:spatial_hadamard_interface}
\end{figure}

Fig.~\ref{fig:spatial_hadamard_interface} shows the results for this reduced noise model.
From these plots we can extract the effective distances: matching-based decoders achieve $d_{\mathrm{eff}} = 2\lfloor k/2 \rfloor + 1$ with all flag configurations [panels (a) and (b)], while the Tesseract decoder achieves $d_{\mathrm{eff}} = 2k+1$ with ``partial'' and ``all'' flags [panel (c)].

\subsection{Decoder Runtime Comparison}
\label{app:decoder_runtime}

\begin{figure}[h]
    \centering
    \includegraphics[width=\columnwidth]{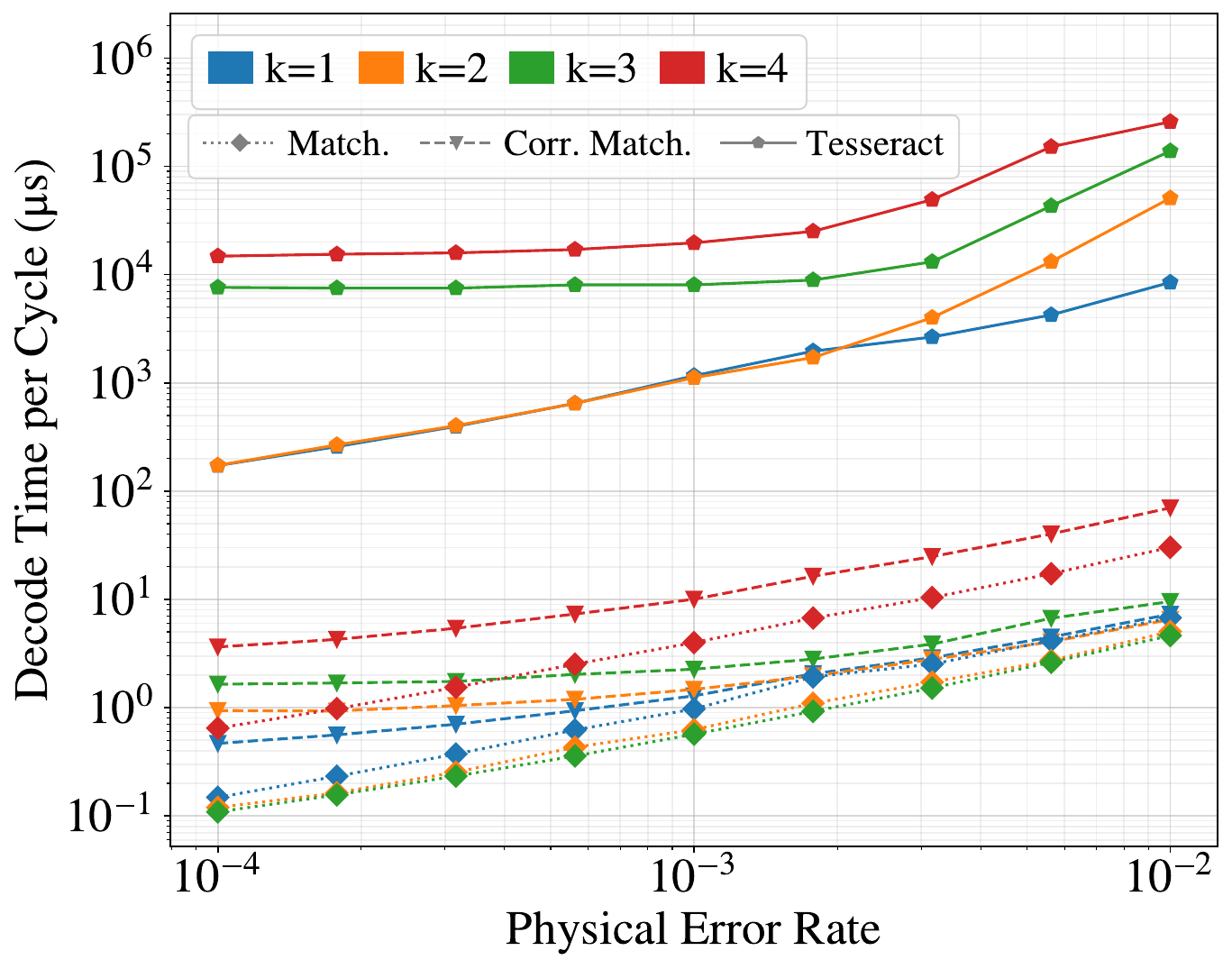}
    \caption{\textbf{Decode time per cycle} versus physical error rate for the \textbf{spatial Hadamard} circuit with the full noise model, for different decoders and code distances.
    The Tesseract decoder (pentagons, solid lines) is typically $10^3$--$10^4$ times slower than matching-based decoders (diamonds/triangles, dotted/dashed lines) across all error rates and distances.
    }
    \label{fig:decoder_runtime}
\end{figure}

While the Tesseract decoder achieves the full circuit-level distance, it comes at a significant computational cost.
Fig.~\ref{fig:decoder_runtime} compares the decode time per cycle for the three decoders across different physical error rates and code distances.
The Tesseract decoder is typically $10^3$--$10^4$ times slower than matching-based decoders, making it impractical for real-time decoding in many applications.
This runtime tradeoff must be weighed against the distance advantage, though as noted in the main text, the distance reduction for matching decoders is limited to the two-dimensional interface surface.

\bibliography{references}

\end{document}